\documentclass[published]{JHEP3} 
\JHEP{01(2004)070}
\received{December 2, 2003}
\accepted{January 30, 2004}
\keywords{AdS-CFT and dS-CFT Correspondence, Cosmology of Theories beyond the SM}

\skip\footins = 1\bigskipamount plus 2pt minus 4pt

\usepackage{epsfig}
\usepackage{amsmath}

\allowdisplaybreaks

\title{Inflationary perturbations from deformed CFT}

\author{Jan Pieter van der Schaar\\ 
CERN Theory Division\\ 
CH-1211 Geneva 23, Switzerland\\
E-mail: \email{Jan.Pieter.van.der.Schaar@cern.ch}}

\abstract{We present a new method to calculate the spectrum of
  (slow-roll) inflationary perturbations, inspired by the conjectured
  dS/CFT correspondence. We show how the standard result for the
  spectrum of inflationary perturbations can be obtained from deformed
  CFT correlators, whose behavior is determined by the Callan-Symanzik
  equation. We discuss the possible advantages of this approach and
  end with some comments on the role of holography in dS/CFT and its
  relation to the universal nature of the spectrum of inflationary
  perturbations.}

\begin{document}

\section{Introduction}

Observational evidence has completely revised our ideas on the history
and future fate of our universe. As long suspected and recently shown
to be in excellent agreement with the Cosmic Microwave Background
(CMB) observations~\cite{WMAP}, the early universe probably
experienced a brief phase of slow-roll inflation, corresponding to
accelerated expansion close to a de Sitter spacetime.  More
surprisingly, supernovae observations~\cite{SN} suggest that our
universe is currently also experiencing a phase of accelerated
expansion, presumably again described by an approximate de Sitter
phase. These phenomenological considerations partly explain the
current surge of theoretical interest in de Sitter spacetime. On the
other hand there have always been strong theoretical motivations to
study de Sitter space.  In particular, since de Sitter spacetime can
be considered the most symmetric cosmological equivalent of a black
hole, it seems to be a nice arena to study some of the confusing
properties having to do with the presence of event horizons.

String theory has made some progress in understanding certain aspects
of black hole physics. An extremely powerful idea that seems to be
concretely realized in string theory is the holographic
principle~\cite{HP}. In the context of Anti-de Sitter gravity the
holographic principle is realized through the AdS/CFT
correspondence~\cite{AdS/CFT, AdS/CFT-rev}.  In a leap of faith
attempts have been made to generalize the AdS/CFT correspondence to
apply it in the context of de Sitter space, leading to the dS/CFT
conjecture~\cite{dS/CFT}. Just as the radial coordinate in AdS space
is related to the scale parameter in the dual lorentzian CFT, in
dS/CFT the timelike coordinate of de Sitter space should be related to
the scale parameter in an euclidean CFT. This also suggests that
(homogeneous) departures from AdS or dS will correspond to
renormalization group (RG) flows in the corresponding deformed
CFT's. However, the absence of supersymmetry, a concrete string theory
example and more fundamental objections~\cite{susskind}, raise severe
doubts on the correctness and usefulness of the dS/CFT
proposal. Nevertheless, from a pragmatic point of view it seems that
some problems in four-dimensional de Sitter space can certainly be
addressed using a general three-dimensional CFT description, as
conjectured by the dS/CFT correspondence but possibly just a result of
the symmetries of the de Sitter background. From that more modest
point of view a CFT approach would just correspond to a different
perspective on some problems in (approximate) de Sitter space, and the
CFT approach would in principle not be able to teach us anything
beyond what was known already. Nevertheless we think this could still
be an interesting and worthwhile enterprise, hopefully giving us at
least some qualitative hints on some of the properties of quantum
gravity in de Sitter spacetime.

As already mentioned, an important point is that homogeneous
deformations away from de Sitter spacetime, in general corresponding
to Friedmann-Robertson-Lemaitre-Walker (FRLW) cosmologies but for
small deformations giving rise to inflationary cosmologies, are
interpreted as departures from perfect scale invariance in the
euclidean CFT. This would then give rise to renormalization group flow
away from the conformal fixed point~\cite{Inflation-dS/CFT, BDM}, a
situation that has been particularly well-studied and understood in
the AdS/CFT context~\cite{PS}--\cite{BFS}.  Since approximate de
Sitter spaces seem to play an important role in the history of our
universe, a holographic RG-flow description could have direct
phenomenological applications. In particular the early inflationary
phase and its production of small perturbations that are the
primordial seeds of the observed CMB, and therefore of large scale
structures in our universe, would be interesting examples to study
from a CFT point of view and is the focus of this article.  One strong
reason to suspect that this should be possible is the approximate
scale invariant nature of the primordial spectrum of density
perturbations~\cite{CMB, CMB-rev, CMB-book}.  This possibility was
first anticipated in~\cite{LvdSL}, and shown to work for massless
perturbations in~\cite{Maldacena}. We will describe the more general
(massive) case and make full use of the holographic RG-equations, that
for the de Sitter case were only determined in all detail very
recently~\cite{Larsen-McNees}. As we will explain such an approach
sheds new light on the universal nature of the spectrum of
inflationary perturbations and the related question of the effect of
trans-Planckian physics on the primordial CMB spectrum, a subject that
has attracted a lot of attention recently~\cite{TP-KKLS}. We stress
that the dS/CFT correspondence is not probing the static patch
describing a single bulk Hubble volume and discuss the meaning of
holography in dS/CFT.  We suggest that the main implication of
holography in cases where the holographic coordinate is timelike
should be universality.

The structure of this paper is as follows. In section 2 we briefly
review the dS/CFT correspondence and the interpretation of broken
scale invariance in terms of holographic RG-flow. We then move on to
present a minimal review of the calculation of the spectrum of
inflationary density perturbations.  After that we show that the
spectrum of inflationary perturbations can be derived starting from
CFT correlators and introducing small deformations satisfying the
Callan-Symanzik equation.  We devote a separate section to a
discussion on the holographic interpretation of the primordial CMB
calculation in terms of CFT correlators.  There we also discuss the
relation between the deformed CFT approach and the universal nature of
the spectrum of inflationary perturbations.
  
\section{dS/CFT and holographic RG-flow}

We give a short review of dS/CFT and the holographic correspondence
between cosmological evolution and renormalization group flow. For
similar, more detailed discussions we refer to~\cite{LvdSL,dS-HRG,NO}.

\subsection{Some dS/CFT preliminaries}

De Sitter space can be obtained from Anti-de Sitter by performing a
so-called double Wick rotation. Starting from the AdS metric in
stereographic coordinates one performs the following transformations
\begin{equation}
\tau \rightarrow i x_{d-1} \, , 
\qquad 
\lambda \rightarrow -i t \, , 
\qquad 
h \rightarrow i H \, , 
\label{AdSdS}
\end{equation}
to obtain de Sitter space in planar, or inflationary, coordinates
(covering only half of de Sitter space)
\begin{equation}
ds^2 = -dt^2 + e^{2Ht} dx_d^2 \, . 
\label{dsinflat}
\end{equation}
Introducing the scale factor $a=e^{Ht}$ and defining the following
coordinate
\begin{equation}
\mu = Ha \, , 
\label{dSscale}
\end{equation}
which has the dimension of energy, we obtain the following
parametrization of $(d+1)$-dimensional de Sitter
\begin{equation}
ds^2 = H^{-2} \left[ - \mu^{-2} d\mu^2 + \mu^2 dx_d^2 \right] . 
\label{dSholo}
\end{equation}
In dS/CFT the parameter $\mu$ will again have the natural
interpretation of the energy scale in a dual CFT description. The
boundaries of the AdS and dS metrics in this ``holographic''
parametrization are obtained in the limit $\mu = u \rightarrow
\infty$. We also note that for de Sitter this parametrization is
related to a de Sitter parametrization using conformal time $\eta$,
which is equal to $\eta = \pm 1 / \mu$ and therefore naturally
interpreted as the length scale in a dual field theory. This
corresponds to the de Sitter analogue of the Poincar{\'e} coordinates
in AdS. De Sitter in conformal time gives rise to the conformally flat
metric
\begin{equation}
ds^2 = (H \eta)^{-2} \left[ - d\eta^2 + dx_d^2 \right] , 
\label{dSconformal}
\end{equation}
where the infinite future corresponds to $\eta=0$ and the infinite
past to $\eta \rightarrow \infty$.

As in AdS/CFT we would now like to introduce perturbations and the
easiest thing to consider is adding scalar fields to de Sitter. As
$t\rightarrow \infty$ the homogeneous part of the scalar field is
dominant, giving us the following asymptotic behavior of the metric
and scalar field
\begin{eqnarray}
ds^2 &\sim& -dt^2 + e^{2Ht} \hat{g}_{ij} dx^i dx^j \, , 
\nonumber \\
\phi &\sim& e^{\lambda Ht} \, \hat{\phi} \, , 
\label{asdsphi}
\end{eqnarray}
where $\lambda_{\pm} = -d/2 \pm \sqrt{d^2/4 - m^2/H^2}$~\cite{dS/CFT}.
Scale invariance demands that under a time translation $Ht \rightarrow
Ht+ \delta$ and $x_i \rightarrow \left( e^{\delta} \right)^{-1} x_i$,
the full asymptotic solution remains invariant. This is ensured if the
(``boundary''\footnote{The ``boundary'' is defined as the spacelike
  section at $t \rightarrow \infty$.})  fields $\hat{g}_{ij}$ and
$\hat{\phi}$ transform as follows
\begin{equation}
\hat{g}_{ij} \rightarrow \left( e^\delta \right)^{0} \hat{g}_{ij} \,, 
\qquad 
\hat{\phi} \rightarrow \left( e^\delta \right)^{-\lambda} \hat{\phi} \, . 
\label{bftransform} 
\end{equation}
In a holographic dual field theory description these boundary fields
are supposed to be sources of dual operators in the field theory. The
metric boundary field $\hat{g}_{ij}$ is the natural source of the
energy momentum tensor in a dual field theory
\begin{equation}
{\cal S}_{\hat{g}} \propto \int d^d x \, \hat{g}_{ij} \, {\cal T}^{ij} \, , 
\label{EMTsource}
\end{equation} 
whereas the scalar boundary fields should correspond to sources of
(gauge invariant) operators in a holographic dual
\begin{equation}
{\cal S}_{\hat{\phi}} \propto \int d^d x \, \hat{\phi} \, {\cal O} \, . 
\label{Osource}
\end{equation} 
From the scaling transformations of the boundary fields and the total
scale invariance of the holographically dual lagrangian we therefore
find the following scaling dimensions ${\cal O} \sim \left( e^\delta
\right)^\Delta {\cal O}$ of the dual operators
\begin{equation}
\Delta_{{\cal T}_{ij}} = d \, , 
\qquad 
\Delta_{\cal O} = \lambda + d \, . 
\label{scalingdimop}
\end{equation}
In case one is interested in (homogeneous) deformations of the CFT,
typically breaking conformal invariance and leading to RG-flow, the
behavior is determined by the scaling dimension of the operator
deforming the theory.  To remind ourselves: marginal operators
correspond to scaling dimensions equal to $d$, so indeed the energy
momentum tensor corresponds to a marginal operator, as it
should. Scaling dimensions $\Delta > d$ correspond to irrelevant
non-renormalizable perturbations, whereas operators with $\Delta < d$
are relevant normalizable operators. Relevant operators will take us
away from the UV fixed point.

Formally the AdS/CFT correspondence can be summarized by the
mathematical statement that the partition function of a dual CFT is
given by the (semi-classical) wavefunction of quantum gravity on AdS
space~\cite{AdS/CFT-def}
\begin{equation}
{\cal Z}_{\rm CFT} = \Psi_{AdS} \approx e^{i\, S_{AdS}} \, . 
\label{ads-cft}
\end{equation}
To make sense out of this expression one typically Wick rotates to
euclidean AdS, giving $\exp{-S_{EAdS}}$ for the euclidean partition
function.  To explicitly relate boundary correlation functions to bulk
quantities we treat the boundary fields $\hat{\phi}$ as sources for
CFT operators and, as usual, functionally differentiate with respect
to the sources, setting the sources to zero afterward. From the bulk
perspective this means that in the semiclassical limit we evaluate the
bulk action on a (normalizable) solution to the equations of motion
with suitable boundary conditions giving the dependence of the bulk
action on $\hat{\phi}$.
\begin{equation}
\left< {\cal O}_1 \ldots {\cal O}_n \right> = \left. { \delta^n {\cal Z}_{\rm CFT} 
\over \delta \hat{\phi}_1 \ldots \delta \hat{\phi}_n} \right|_{\hat{\phi}=0} 
= \left. { \delta^n \Psi_{AdS}[\hat{\phi}] 
\over \delta \hat{\phi}_1 \ldots \delta \hat{\phi}_n} \right|_{\hat{\phi}=0}, 
\label{corrfns}
\end{equation}
where $\bar{\phi}$ is the source of the CFT operator ${\cal O}$.  Any
singularities that arise in the action should be treated just as if
one is renormalizing a field theory and will typically lead to
(holographic) RG-flow~\cite{BVV,BFS}.  As already alluded to this
RG-flow can be understood as being described by a deformation with a
local operator
\begin{equation}
S_{\rm CFT} \rightarrow S_{\rm CFT} + g \int d^d x {\cal O}(x) \, , 
\label{cftdeform}
\end{equation}
which by means of the correspondence~(\ref{Osource}) suggests a
relation to homogeneous bulk gravitational solutions $\phi \sim g$.
Deforming the CFT with an operator of dimension $\Delta$ implies the
lagrangian deformation scales as $\Delta -d =
\lambda$~(\ref{scalingdimop}).  This coincides with the asymptotic
behavior of (homogeneous) scalar fields~(\ref{asdsphi}).  Relevant
deformations vanish in the UV to be consistent with the presence of a
conformal fixed point, whereas irrelevant deformations blow up in the
UV.  From the bulk perspective this is just saying that asymptotic de
Sitter space is only consistent with decaying behavior of the
(homogeneous) scalar field.  When the scalar field blows up de Sitter
space is not the asymptotic solution.

Moving from AdS to dS we will assume that the same recipe applies for
dS/CFT, i.e.\
\begin{equation}
{\cal Z}_{\rm CFT} = \Psi_{dS} \approx e^{i\, S_{dS}} \, , 
\label{ds-cft}
\end{equation}
where the CFT in this case is already defined in euclidean space and
the right hand side corresponds to the wavefunction of quantum gravity
in de Sitter space, which can now honestly be interpreted as the
``wavefunction of the universe''. As in AdS the de Sitter action will
generically diverge when approaching the ``boundary'' in the infinite
future. Just as in field theory, the (local) divergences can be
canceled by counterterms leading to a renormalized action satisfying
RG equations. For de Sitter this holographic renormalization procedure
has only recently been carefully worked out~\cite{Larsen-McNees}.

\subsection{Homogeneous monotonic flow and its holographic description}

Consider $(d+1)$-dimensional Einstein gravity coupled to scalars
$\phi^I$ \emph{via}
\begin{equation}
{\cal L}={1\over 2 \kappa^2} R - {1\over 2} G_{IJ} \partial_\mu \phi^I
\, \partial^\mu \phi^J - V(\phi^I) \,,
\label{lagrangian}
\end{equation}
where $\kappa^2= 8\pi G= 1/{M_p}^{(d-1)}$ and $G_{IJ}$ is the metric
on the moduli space of the scalars $\phi^I$, which we will assume to
be independent of the scalars $\phi^I$ for the time being. It could
perhaps be interesting to relax this assumption which would involve
introducing covariant derivatives with respect to the moduli space
metric.

\pagebreak[3] 

The potential $V(\phi^I)$ is assumed to have de Sitter extrema. Using
the standard spatially flat FRW metric
\begin{equation}
ds^2= -dt^2 + a(t)^2 dx_i dx^i\,,
\label{FRWansatz}
\end{equation}
with $i$ running from $1$ to $d$ and assuming spatial isotropy
$\vec{\nabla} \phi^I=0$, the equations of motion are given by
\begin{eqnarray}
\ddot{\phi}^I + d\, H \dot{\phi}^I + G^{IJ} {dV \over d\phi^J} &=& 0 \,, 
\label{scalareom} \\
H^2 &=& {2 \over d(d-1)} \kappa^2 \rho \,, 
\label{H2eom} \\
\dot{H} &=& -{1\over (d-1)} \kappa^2 \left( \rho + p \right) ,
\label{Hdoteom}
\end{eqnarray}
where the Hubble parameter $H$ is $H\equiv \dot a/a$, the dot
representing a derivative with respect to the coordinate time $t$, and
the density $\rho$ and pressure $p$ are given by
\begin{eqnarray}
\rho &=& {1\over 2} G_{IJ} \dot{\phi}^I \dot{\phi}^J + V(\phi^I)\,,
\nonumber \\
p &=& {1\over 2} G_{IJ} \dot{\phi}^I \dot{\phi}^J - V(\phi^I)\,.
\label{rhoand}
\end{eqnarray} 
The equations of motion give
\begin{equation} 
d (\omega+1) = -{ 2 \dot{H} \over H^2} \,, 
\label{eos}
\end{equation}
for the equation of state parameter $\omega \equiv p / \rho$. Also
recall that one of the equations of motion is redundant,
e.g.~(\ref{scalareom}) and~(\ref{H2eom}) imply~(\ref{Hdoteom}). For
completeness let us also separately give the expression for the Ricci
scalar $R$
\begin{equation}
R\equiv g_{\mu\nu} R^{\mu\nu} = {2\kappa^2 \over d-1} (\rho - dp) = 2\kappa^2 
\left[ \left({ d+1 \over d-1} \right) V(\phi^I) - {1\over 2} G_{IJ} 
\dot{\phi}^I \dot{\phi}^J \right] .
\label{curveR}
\end{equation}
Note that for constant fields $\phi^I$ we find
\begin{equation}
R=\left({d+1 \over d-1}\right) \, 2\kappa^2 V(\phi^I) = 
d(d+1) H^2 = \left({d+1 \over d-1}\right) \Lambda \, , 
\label{RLH}
\end{equation}
where we used that $\Lambda \equiv 2\kappa^2 V(\phi^I)$.

From now on we will restrict ourselves to four bulk spacetime
dimensions ($d=3$) because we will be interested in calculating the
primordial CMB spectrum. We will also restrict our attention to a
single scalar. The multi-scalar generalization should be rather
straightforward.  From the previous section it should be clear that we
can interpret the scale factor $a$ as proportional to the (deformed)
CFT scale parameter $\mu \propto a$~(\ref{dSscale}) and every scalar
$\phi$ as a coupling constant $g=\kappa \phi$ deforming the
CFT~(\ref{Osource}).  This means we can naturally define the beta
functions
\begin{equation}
\beta \equiv {dg \over dlog \mu} = {\kappa \over H} \dot{\phi} \,  
\label{bfns}
\end{equation}
which by virtue of the FRW equation~(\ref{Hdoteom}) have to satisfy
\begin{equation}
\beta \left( \beta + {2 \over \kappa H} {dH \over d\phi} \right)=0 \, .
\label{Hbeta1}
\end{equation} 
Under the natural assumption of monotonic scalar field flow we derive
\begin{equation}
\beta = -{2 \over \kappa H} {dH \over d\phi} \, .
\label{Hbeta}
\end{equation}
Monotonicity implies that we will not consider scalar field flows that
are oscillating in time. This means we will only consider masses $m^2
< H^2$, which is the relevant regime anyway for the spectrum of
inflationary perturbations that we will be interested in.

Under this monotonicity assumption we can relate the equation of state
parameter~(\ref{eos}) and the beta function~(\ref{Hbeta}) as follows
\begin{equation}
3(\omega +1) = {4 \over \kappa^2} H^{-2} \left({d H \over 
d\phi}\right)^2 =  \beta^2 \, .
\label{omegabeta}
\end{equation}
Recall that for constant equations of state, i.e.\ matter ($\omega=0$)
or radiation ($\omega=1/3$), the dependence of the density on the
scale factor would be $\rho \propto a^{-3(\omega+1)}=a^{-\beta^2}$.
To decide whether the expansion of the universe is accelerating or
decelerating we calculate
\begin{equation}
{\ddot{a} \over a} = \dot{H} + H^2 = H^2 \left( 1-{1 \over 2} \beta^2 \right),
\label{acdec}
\end{equation}
so whenever $\beta^2 < 2$ ($\omega < -1/3$) the expansion of the
universe is accelerating.

Let us also define a function $\lambda$ that will reproduce the
conformal scaling dimensions, derived as follows from the
beta-functions
\begin{equation}
\lambda \equiv {1\over \kappa} {d \beta \over d \phi} \, .
\label{sdmatrix}
\end{equation}
This implies that near a zero of $\beta$ we can approximately write
\begin{equation}
\lambda = {1 \over \kappa} \left.{d\beta \over d\phi} 
\right|_{\beta =0} \approx -{2 \over \kappa^2 H} \left.{d^2 H \over d\phi 
d\phi} \right|_{\beta =0} \, . 
\label{sdapprox}
\end{equation}
One can show that near a zero of $\beta$, so the extrema of the
potential, one can relate this expression to the mass$^2$ in the
following way
\begin{equation}
m^2 = -H^2 ( 3 \lambda + \lambda^2 ) \, , \label{M2sdmatrix}
\end{equation}
which just reduces to the standard equations relating scaling
dimensions to masses.

For completeness let us also introduce the conjectured holographic
c-function~\cite{Inflation-dS/CFT} given~by
\begin{equation}
c \equiv {1 \over \kappa^2 H^2}
\label{holc}
\end{equation}
and is strictly decreasing from the UV to the IR fixed point because
$\dot{H} < 0$ when the Null Energy Condition is satisfied $\omega \geq
-1$.  One can relate the derivative of the holographic c-function with
respect to the scalar fields to the beta-function as follows
\begin{equation}
{d\ln{c} \over \kappa d\phi} = \beta \, .
\label{cbeta}
\end{equation}
Instead taking the derivative with respect to the logarithmic energy
scale (the logarithm of the cosmological scale factor $a$, which is
related to the number of e-folds $N$) we obtain
\begin{equation}
{d\ln{c} \over d\ln{a}} = {d\ln{c} \over dN} = \beta^2 \, .
\label{cbeta2}
\end{equation}
So this implies we can write
\begin{equation}
c(N_f)=c(N_i) \exp{\int_{N_i}^{N_f} \beta^2 dN} \, .
\label{cintN}
\end{equation}
Some of these expressions will turn out to be useful later. Note that
the entropy in (the static or causal patch of) de Sitter space is
given by the area of a single Hubble volume, which is proportional to
the holographic c-function
\begin{equation}
S = {Area \over 4\pi G} = {2\pi \over \kappa^2 H^2} = 2\pi \, c \, .
\label{S-c}
\end{equation} 

After these preliminaries we are now ready to move on to discuss
(inhomogeneous) perturbations in de Sitter space.

\section{Inflationary perturbations}

We will present a rather minimal review of inflationary (density)
perturbations, concentrating on the final result. More thorough
introductions can be found in many places, we would especially like to
mention~\cite{CMB-rev, CMB-book, Maldacena}. In particular, bulk
approaches based on a Hamilton-Jacobi setup~\cite{SB, BVV,
  Larsen-McNees} make the appearance of the holographic RG-equations
and the possibility to derive the inflationary perturbations from
deformed CFT correlators very apparent and natural.

\subsection{De Sitter scalar perturbations}

Abandoning homogeneity of the scalar field and neglecting
gravitational back reaction one should solve the following equation
\begin{equation}
\ddot{\delta\phi} + 3 H \dot{\delta\phi} - {1\over a^2} \vec{\nabla}^2 
\delta\phi + {dV \over d\delta\phi} = 0 \, , 
\label{inhomosf} 
\end{equation}
We will be expanding around an extremum of the potential, so we can
approximate ${dV \over d\phi} \approx m^2 \delta\phi$. To comply with
the standard literature we define conformal time $\eta$ and redefine
the scalar field as follows
\begin{equation}
\eta \equiv -{1 \over Ha} = -{1\over \dot{a}} \,, 
\qquad 
\varphi \equiv a \delta\phi \, , 
\label{redefine}
\end{equation}
where $a$ is the scale factor and $H={\dot{a} \over a}$. Note that the
conformal time coordinate equals $1/\mu$, where $\mu$ can be
interpreted as the natural scale parameter in a dual CFT. This enables
us to write the scalar equation as
\begin{equation}
\varphi^{\prime\prime} + \left( {m^2 \over H^2} -2 \right) {1\over \eta^2}
\varphi - \vec{\nabla}^2 \varphi = 0 \, , 
\label{conftimeq}
\end{equation}
where double primes denote the second derivative with respect to the
conformal time $\eta$. Fourier expanding in the spatial directions
\begin{equation}
\varphi(\eta, \vec{x}) = \int d^3k \varphi_k({\eta}) e^{i\vec{k} \cdot 
\vec{x}} \, , 
\label{fourier}
\end{equation}
we finally obtain the following equation for the time dependence of
the Fourier components~$\varphi({\eta})$
\begin{equation}
\varphi_k^{\prime\prime} + \left( k^2 + {m^2/H^2 -2 \over \eta^2} \right)
\varphi_k =0 \, . 
\label{kmodeq}
\end{equation}
When $m^2=0$ this equation is relatively easy to solve and we get
\begin{equation}
\varphi_k \propto \left( 1 \pm {i \over \sqrt{k^2} \eta} \right) 
e^{\pm i \sqrt{k^2} \eta} \, . 
\label{m2=0sol}
\end{equation}
For arbitrary $m^2$ it is more convenient to first proceed as follows.
Define $\chi \equiv k \eta = k/\mu$ and $\sqrt{\chi} \psi(\chi) \equiv
\varphi(\chi)$. This transforms equation~(\ref{kmodeq}) into
\begin{equation}
\psi_k^{\prime\prime} + {1\over \chi} \psi_k^{\prime} + \left( 1 - 
{n^2 \over \chi^2} \right) \psi_k =0 \, ,
\label{besseleq}
\end{equation}
where the primes now stand for derivatives with respect to $\chi$ and
$n^2 \equiv {9\over 4} - {m^2 \over H^2}$. This is recognized as a
Bessel equation with non-integer coefficient $n^2$.  Note that for
$m^2 < 9H^2/4$ the coefficient $n$ is real, whereas for $m^2 > 9H^2/4$
it is imaginary. This Bessel equation is solved by the series
expansion
\begin{equation}
\psi_k (\chi)= \chi^n \, \left[ 1- {1\over n+1} \left({\chi \over 2}\right)^2 
+ {1\over (n+1)(n+2)} {1\over 2!} \left({\chi \over 2}\right)^4 
- \cdots \right] . 
\label{besselsol}
\end{equation}
Defining the Bessel functions as
\begin{equation}
{\cal J}_n (\chi) = \sum_{r=0}^{\infty} \, {(-1)^r \over r! \, \Gamma(n+r+1)}
\left({\chi \over 2}\right)^{n+2r} \, , 
\label{besfndef}
\end{equation}
and the solution would read $\psi(\chi)= n! \, 2^n \, {\cal J}_n
(x)$. Note that for half-integral $n$ the Bessel functions can be
simply expressed in terms of trigonometric functions. Starting from
${\cal J}_{1/2} (\chi) = ({2 / \pi \chi})^{1/2} \sin \chi$ and ${\cal
  J}_{-1/2} (\chi) = ({2 / \pi\chi})^{1/2} \cos \chi$ one can use
recursion relations to show that $\psi_k (\chi) = \sqrt{2\over \pi
  \chi} ( 1 \pm {i \over \chi} ) \exp{\pm i \chi}$,
reproducing the solution for $m^2=0$ ($n=\pm 3/2$).

Introducing a second independent solution to~(\ref{besseleq})
\begin{equation}
{\cal Y}_n ={ \cos n\pi \, {\cal J}_n - {\cal J}_{-n} \over
\sin n\pi} \, . \label{indepsol}
\end{equation}
we can construct the independent solutions known as the Hankel
functions
\begin{eqnarray}
{\cal H}_n^{(1)} &=& {\cal J}_n  + i \, {\cal Y}_n \, , 
\label{hankel1} \\
{\cal H}_n^{(2)} &=& {\cal J}_n  - i \, {\cal Y}_n  \, . 
\label{hankel2}
\end{eqnarray}
The most general solution can then be written in terms of the Hankel 
functions
\begin{equation}
\psi_k (\chi) = A(k)\, {\cal H}_n^{(1)} (\chi) + B(k)\, {\cal H}_n^{(2)} 
(\chi) \, ,\label{gensol}
\end{equation}
where $A(k)$ and $B(k)$ are arbitrary constants determined by the
boundary and normalization conditions. Normalization should be carried
out with respect to the standard (conserved) Klein-Gordon inner
product
\begin{equation}
i \int d\vec{x} \sqrt{\det g_{ij}} \, \phi(\vec{x},t)^\star 
\overleftrightarrow{\partial_t} \phi(\vec{x},t) = 1 \, , \label{KGnorm}
\end{equation}
which in momentum space reduces to the following Wronskian
\begin{equation}
 (-i) |k| \chi \left[\psi_k^\star \partial_\chi \psi_k - (\partial_\chi 
\psi_k^\star)\psi_k \right] = 1 \, . \label{knorm}
\end{equation}
To determine the constant $A(k)$ that normalizes the solution it is
convenient to first determine the asymptotic behaviour as $|\chi| \gg
1$ of the Hankel functions, which up to constant phases behave as
follows
\begin{eqnarray}
{\cal H}_n^{(1)}(\chi) &\sim& \sqrt{2\over \pi \chi} e^{i \chi} 
\, , \label{limh1} \\
{\cal H}_n^{(2)}(\chi) &\sim& \sqrt{2\over \pi \chi} e^{-i\chi} 
\, . \label{limh2}
\end{eqnarray}
Note that $|\chi| \gg 1$ corresponds to $a \ll 1$, so it corresponds
to the scalar field behaviour in the far past of de Sitter space. The
limiting behaviour in the far future of de Sitter space corresponds to
taking the limit $|\chi| \ll 1$, which in terms of the natural
holographic parameter $\mu$ corresponds to $k \ll \mu$, and
immediately follows from the expansion~(\ref{besselsol})
\begin{equation}
\lim_{|\chi| \rightarrow 0} \psi_k (\chi) \sim \chi^n \, , \label{xolimit}
\end{equation}
which indeed corresponds to the expected scaling behaviour of the
scalar field in the far future. We observe that the solutions start
out as ordinary oscillating plane waves but at some point stop
oscillating (if $n$ is real) with slowly decaying, almost constant,
amplitudes. This means that quantum fluctuations, starting out as
ordinary plane waves, will at some point stop oscillating with their
amplitudes almost frozen. This is the mechanism for the production of
classical, stochastic, scalar fluctuations in de Sitter
space. Returning to the normalization of the solutions, consider the
solution $\psi_k (\chi) = A(k) {\cal H}_n^{(1)}(\chi)$ in the limit
$|\chi| \gg 1$, using expression~(\ref{limh1}), and substitute it
into~(\ref{knorm}).  Note that we have imposed a boundary condition by
setting $B(k)=0$, which is the natural choice from the QFT point of
view, where this would correspond to selecting the standard euclidean,
Bunch-Davies vacuum. Other choices have recently attracted a lot of
attention because of their potential connection to trans-Planckian
physics~\cite{Alpha}. Proceeding, one then finds for $A(k)$
\begin{equation}
A(k) = \pm \sqrt{\pi \over 4k} \, . \label{normAk}
\end{equation}
For later purposes it is useful to determine the $H$, $a$ and $k$-dependence 
of the solutions in the limit $|\chi| \ll 1$ (or $a \gg k/H$).
Using~(\ref{xolimit}) and realizing that the variable 
$\chi = -k/Ha$ we find for the $H$, $a$ and $k$ dependence 
\begin{equation}
\delta\phi_k = i {H \over k} \chi^{3\over 2} A(k) \psi_k \propto 
H^{-(n+{1\over 2})} \, k^n \, a^{-(n+{3 \over 2})} \, . \label{phikdep}
\end{equation}
The dominant behavior when $a \gg k/H$ ($\chi \ll 1$) of a general
solution will be given by the $n_- = - \sqrt{{9\over 4} - {m^2 \over
    H^2}}$ mode.  Because $n_- = -3/2 - \lambda_+$ we find the
following dominant (scaling) behavior as $a \gg k/H$ in terms of
$\lambda_+$
\begin{equation}
\lim_{a \rightarrow \infty} \delta\phi_k \propto H^{\lambda_+ + 1} \, 
k^{- {3\over 2}- \lambda_+} \, a^{\lambda_+} \, . \label{domlplus}
\end{equation}  
From this expression it is clear that the amplitude of the mode will
not change much as time evolves from the time the mode is ``frozen'',
at least as long as $|\lambda_+| \ll 1$ implying $m^2 \ll H^2$.  This
moment can be estimated to be around $|\chi|=1$ or when $a=a_*=k/H$,
when the physical wave number equals the Hubble parameter.

\subsection{The CMB power spectrum and spectral indices}

One can view the spectrum of scalar field fluctuations in (pure) de
Sitter space as the source for scalar curvature perturbations during
inflation.  The mean square scalar field fluctuations in Fourier space
are given by $|\delta\phi_k|^2$. The power spectrum can be written as
\begin{equation}
{\cal P}_{\delta\phi} = {k^3 \over 2\pi^2} \, |\delta\phi_k|^2 \, . 
\label{sfpower}
\end{equation}
This can be traced back to the following general definition of a power
spectrum
\begin{equation}
\left< g_{\vec{k}} g_{\vec{k}^\prime} \right> = \delta^{(3)}\left(\vec{k}-
\vec{k}^\prime\right) |g_{k}|^2 \equiv \delta^{(3)}\left(\vec{k}-
\vec{k}^\prime\right) {2\pi^2 \over k^3} {\cal P}_g(k) \, , 
\label{powerdef}
\end{equation}
where $g_k(t)$ corresponds to the Fourier transform of some quantity
$g(\vec{x},t)$. This definition immediately gives~(\ref{sfpower}) and
also implies the relation
\begin{equation}
\left< g^2(\vec{x},t) \right> = \int {d^3k \over (2\pi)^3}
|g_k|^2 = \int d\ln k \, {\cal P}_g(k) \, . \label{intpower}
\end{equation}
At physical scales $p=k/a$ bigger than the Hubble scale $H$ the mean
square fluctuations behave as ordinary quantum fluctuations in flat
space, which become negligible as one increases the length
scale. However, as already explained, at physical scales $p$ smaller
than the Hubble scale $H$ the field no longer oscillates, leading to
large length scale fluctuations.  Being interested in those scales, we
use the expression for $\phi_k$ as $a\rightarrow \infty$ and obtain
for the power spectrum, neglecting constants,
\begin{equation}
{\cal P}_{\delta\phi} \propto H^{2+2\lambda_+}\, k^{-2\lambda_+} \, 
a^{2\lambda_+} \, . \label{sfpowerlimit}
\end{equation}
The massless limit corresponds to taking $\lambda_+=0$.

\pagebreak[3] 

In pure de Sitter space this scalar field source will not lead to
scalar curvature perturbations because the scalar field fluctuations
are perfectly decoupled in that case. Another way of putting this is
that in pure de Sitter these fluctuations are pure gauge. For instance
consider a density perturbation $\delta \rho$, then it is easy to see
that
\begin{equation}
\kappa \delta \rho = \kappa {d \rho \over d \phi} \, \delta \phi =
(-3 H^2 \beta) \delta \phi \, , \label{densitypert}
\end{equation} 
where $\beta$ is the holographic beta-function defined
in~(\ref{bfns}).  Only when there exists a small ``tilt'' away from
pure de Sitter space ($\beta \neq 0$) will the scalar field
fluctuations couple and induce density perturbations. To obtain the
correct physical quantity denoting the scalar curvature perturbations
and their relation to the scalar field fluctuations is not
straightforward and we will just refer to the
literature~\cite{CMB-rev, CMB-book}. We will just quote the standard
result that the scalar curvature perturbation can be described by the
(dimensionless) quantity
\begin{equation}
\delta {\cal R} = {\delta \rho \over p+\rho} = \left( {-3 \kappa \over \beta}
\right)  \delta \phi \, , \label{deltaR}
\end{equation} 
immediately leading to the following relation between the power spectra 
\begin{equation}
{\cal P}_{\cal R} = \beta^{-2} {\cal P}_{\delta\phi} \, . \label{scpower}
\end{equation}
When we will discuss this result from the deformed CFT point of view the 
extra factor of $\beta$ actually appears rather naturally. 
Note that the factor of $\beta$, for small $\beta \ll 1$, will kill 
the time (or scale factor) dependence of the perturbations because
\begin{equation}
\beta \approx \lambda \delta\phi \Rightarrow \delta\phi \propto a^\lambda
\Rightarrow \beta \propto \lambda a^\lambda \, . \label{adepbeta}
\end{equation}
Before we go on to discuss the spectral index we should note that to
connect to the standard expressions in the literature involving the
slow-roll parameters $\epsilon_H$ and $\eta_H$ one uses the following
definitions for the slow-roll parameters
\begin{equation}
\epsilon_H \equiv - {\partial \ln H \over \partial \ln a} \, , 
\qquad 
\eta_H \equiv - {\partial \ln{\partial H \over \partial \phi} \over
\partial \ln a} \, , \label{slowrollH}
\end{equation}
leading to the following relations between the different sets of
parameters~\cite{LvdSL}: $2 \epsilon_H = \beta^2$ and $2 \eta_H =
\beta^2 - 2 \lambda$.

We are now ready to discuss the spectral index of the scalar curvature
perturbation. Since we just concluded that~(\ref{scpower}) is independent
of time we can evaluate this expression at the time it was formed, which was
estimated to be around $|\chi|=1$, so when $a=k/H$. Plugging this into the
expression~(\ref{sfpowerlimit}) we see that the $k$-dependence disappears and
that we are effectively left with a power spectrum of massless perturbations.
So our starting point is the following expression for the power spectrum of 
the scalar density perturbation
\begin{equation}
{\cal P}_{\rm scalar} \propto \left. \left({H \over \beta}\right)^2 \right|_{k=aH},
\label{powspect}
\end{equation}
where we have added explicitly that this expression should be
evaluated at $k=aH$.  The Hubble parameter $H(a)$ and the
beta-function $\beta(a)$ generically are functions of the scale factor
$a$. Different scales $k$ ``freeze'' into the power spectrum at
different scale factors $a(t)$ when the physical momentum $p=k/a=H$ or
equivalently when $k=aH$. This effectively leads to $k$-dependence of
the spectrum because $\beta$ and $H$ are functions of $a(t)$. To
calculate this scale dependence we first relate derivatives with
respect to $a$ to derivatives with respect to $k$
\begin{equation}
{dk \over da} =\left( H + a {dH \over da} \right) = \left( 1-{1\over 2} 
\beta^2 \right) H \, . \label{dkda}
\end{equation}
This implies that 
\begin{equation}
a{d \over da} = \left( 1-{1\over 2}\beta^2 \right) \, k {d\over dk} \, .
\label{adakdk}
\end{equation}
This allows us to calculate the effective $k$-dependence of $H$ by
noting that
\begin{equation}
a{dH \over da} = - {1\over 2} \beta^2 \, H = \left( 1-{1\over 2} \beta^2 
\right) k{dH \over dk} \, . \label{adHda}
\end{equation}
So in the end we find the following equation for the effective
$k$-dependence of $H$
\begin{equation}
{d\, \ln H \over d\, \ln k} = {-{1\over 2}\beta^2 \over 
( 1-{1\over2}\beta^2 )} \, . \label{kdepHeq}
\end{equation}
Under the assumption that $\beta$ is approximately constant
$\beta=\bar{\beta}$ (i.e.\ $k$-independent) and to lowest order in
$\beta$ this equation is easy to solve and gives
\begin{equation}
H(k) \approx k^{-{1\over2} \bar{\beta}^2} \, . \label{Hpowerk}
\end{equation}

The same procedure can be applied to find the effective $k$-dependence
of $\beta$. We first note that
\begin{equation}
a{d\beta \over da} = \lambda \beta =  \left( 1-{1\over 2} \beta^2 
\right) k{d\beta \over dk} \, , \label{adbetada}
\end{equation} 
where we defined $\lambda$ as $\lambda\equiv {d\beta / d\phi}$
(instead of $\lambda\equiv {\beta /\phi}$).  This implies the
following equation for the effective $k$-dependence of $\beta$
\begin{equation}
{d\, \ln \beta \over d\, \ln k} = {\lambda \over \left( 1-{1\over2}\beta^2 
\right)} \, . \label{kdepbeq}
\end{equation}
Again, under the assumption that $\lambda$ is approximately
$k$-independent $\lambda = \bar{\lambda}$ and to lowest order in
$\beta$ we find
\begin{equation}
\beta(k) \approx k^{\bar{\lambda}} \, . \label{bpowerk}
\end{equation}

We are now ready to calculate the spectral index of the power spectrum
given by~(\ref{powspect}). Typically the spectral index is defined as
follows
\begin{equation}
n_s -1 \equiv {d\, \ln{\cal P}_{\rm scalar} \over d\, \ln k} \, . 
\label{indexdef}
\end{equation}
Through this definition we can in fact calculate the spectral index
\emph{exactly}, using the expressions in the previous paragraphs. But
first let us consider the situation where $\beta$ and $\lambda$ are
approximately $k$-independent and concentrate on the power of $k$ in
the power spectrum, which should naturally define the ``true''
spectral index, as we will discuss in a moment. From~(\ref{Hpowerk})
and~(\ref{bpowerk}) we then obtain for the (average) spectral index
\begin{equation}
\bar{n}_s -1 = -\bar{\beta}^2 - 2\bar{\lambda} \, . \label{avpowerspec}
\end{equation} 
Instead using the definition~(\ref{indexdef}) we find
\begin{equation}
n_s -1 \equiv {d\, \ln{\cal P}_{\rm scalar} \over d\, \ln k} = { -\beta^2 - 
2\lambda \over ( 1- {1\over2} \beta^2 )} \, , \label{indexexact}
\end{equation}
which indeed corresponds to the standard expression for the spectral
index in terms of the slow-roll parameters to lowest order in
$\beta^2$; using~(\ref{slowrollH}) we reproduce $n_s -1=-4\epsilon_H +
2\eta_H$.

One should realize that~(\ref{indexdef}) will only reproduce the
\emph{true} spectral index when $n_s$ is $k$-independent. The true
definition of the $k$-dependent spectral index should be $n(k) -1=
{\ln{\cal P}_{\rm scalar} / \ln k}$, as is clear when writing ${\cal
  P}_{\rm scalar} \propto k^{(n(k) -1)}$.  We can find the following
expression for the difference ${\cal E}$ between the true spectral
index and the spectral index calculated with~(\ref{indexdef})
\begin{equation}
{\cal E} \equiv n_s(k) - n(k) = \left( {d\,n \over d\,\ln k} \right) \ln k \, .
\label{errorindex}
\end{equation}
We can use this expression iteratively to express the true spectral
index $n(k)$ in terms of an infinite derivative expansion of the
other, exactly calculable, spectral index defined
through~(\ref{indexdef}) $n_s(k)$
\begin{equation}
n(k) = n_s(k) - {d\,n_s \over d\, \ln k}\, \ln k + {d^2\,n_s \over (d\, \ln 
k)^2}\, (\ln k)^2 - \cdots \, , \label{errorexpand}
\end{equation}
where the signs alternate between odd and even powers of $\ln k$.  The
first derivative of $n_s$ with respect to $\ln k$ can be calculated
exactly and gives
\begin{equation}
{d\,n_s \over d\,\ln k} = -{2 \over \left(1-{1\over 2}\beta^2 \right)^3}
\left( \beta^2 \lambda (1 + \lambda) + \beta \left(1-{1\over 2}\beta^2\right) 
{d^2 \, \beta \over d\phi^2} \right) . \label{dnsexact}
\end{equation}
To lowest order one therefore finds
\begin{equation}
n(k) = n_s(k) + 2 \beta {d^2 \, \beta \over d\phi^2} \, \ln k + 
2\beta^2 \lambda \, \ln k \, . \label{error1st}
\end{equation}
Clearly, when $\beta, \lambda \ll 1$ (slow-roll regime) and when
$\lambda$ is not changing too rapidly or for ranges of scales $k$ that
are not too large, using the definition~(\ref{indexdef}) is not a bad
approximation.

\section{Inflationary perturbations from CFT correlators}

Soon after the dS/CFT correspondence was first conjectured, it was
realized that it could perhaps have interesting applications in the
context of inflation~\cite{Inflation-dS/CFT}. That one can derive the
power spectrum of inflationary perturbations from the CFT was first
anticipated in~\cite{LvdSL}, and was explained for massless scalar
field perturbations in~\cite{Maldacena}. The results described in this
section can also be found in section 6 of the recent
preprint~\cite{Larsen-McNees}, which appeared during the completion of
this paper.

\subsection{Relating correlation functions to expectation values}

Before we can go on to derive the power spectrum of inflationary
perturbations from deformed CFT correlators we need to know how to
relate expectation values in the dS bulk to CFT correlators. This was
explained in~\cite{Maldacena} and we will briefly review that result
here.

CFT vacuum correlation functions can be deduced from the de Sitter
bulk following the same procedure as in AdS; calculating the action
and functionally differentiating with respect to the boundary fields
that act as sources in the CFT~(\ref{corrfns}). Specifically for CFT
2-pt functions of operators ${\cal O}$ with scaling dimension $\Delta$
this implies
\begin{equation}
\left< {\cal O}_\Delta(x) {\cal O}_\Delta(y) \right> = \left. { \delta^2 
\Psi_{dS}[\phi] \over \delta \phi(x) \delta \phi(y)} 
\right|_{\phi=0} . \label{2ptcorfn}
\end{equation}
At the same time it should be clear that the spectrum of scalar field
fluctuations in de Sitter space that we calculated is related to the
(equal time) expectation value $| \phi_k |^2= \int d^3 k' \langle
\phi_{\vec{k}} \phi_{\vec{k}'} \rangle$. A scalar field expectation
value can be expressed, up to a delta function, as a functional
integral of $\phi_k^2$ weighted by the norm of the scalar field wave
functional squared. The scalar field wave functional can be
approximated by the classical action in de Sitter (in the
semi-classical limit). This leads to the following formal expression
for the expectation value of scalar field perturbations as
\begin{equation}
\left< \phi_{\vec{k}} \phi_{-\vec{k}} \right> = \int 
{\cal D}\phi \, \phi_{\vec{k}} \phi_{-\vec{k}} \, \left| \Psi_{dS}[\phi] 
\right|^2 \approx \int {\cal D}\phi \, \phi^2 \, \left| e^{iS_{dS}[\phi]} 
\right|^2  . \label{bulkev}
\end{equation}
From these expressions one can find a direct relation between de
Sitter expectation values in the bulk and CFT 2-pt functions. By
definition~(\ref{2ptcorfn}) one can formally express the wave
functional (or generating functional) $\Psi_{dS}$ in terms of the 2-pt
function and higher order (connected) $n$-point functions
\begin{equation}
\Psi_{dS}[\phi] = e^{{1\over 2} \int d^3k \, d^3k^\prime \, \left<
  {\cal O}_{\vec{k}} {\cal O}_{\vec{k}^\prime} \right> \,
  \phi_{\vec{k}} \phi_{\vec{k}^\prime} + \cdots + {1\over n!}  \int
  d^3k \ldots d^3k^n \, \phi_{\vec{k}} \ldots \phi_{\vec{k}^n} \left<
  {\cal O}_{\vec{k}} \ldots {\cal O}_{\vec{k}^n} \right>} \,
. \label{nptexpand}
\end{equation} 
Plugging this expression into the expression for the expectation
value~(\ref{bulkev}), neglecting the higher order $n$-point functions,
we obtain
\begin{equation}
\left< \phi_k \phi_{-k} \right> = \int {\cal D}\phi \, \phi^2 \, 
e^{\int d^3k \int d^3k^\prime \, \phi_k \phi_{k'} \,  {\rm Re} 
\left< {\cal O}_k {\cal O}_{k'} \right>}  \, . 
\label{expectOO}
\end{equation}
By changing the path integral variable (without changing the measure)
we can isolate the dependence on the 2-pt correlation
function. Defining
\begin{equation}
\tilde{\phi}_{\vec{k}} \equiv i \sqrt{2 {\rm Re} \left< {\cal O}_k 
{\cal O}_{-k} \right>} \, \phi_{\vec{k}} \, , \label{intvartrans}
\end{equation}
we obtain the following expression for the expectation value
\begin{equation}
\left< \phi_k \phi_{-k} \right> = {-1 \over 2 {\rm Re} \left< {\cal O}_k 
{\cal O}_{-k} \right>} \, \int {\cal D}\tilde{\phi} \, \tilde{\phi}^2 \, 
e^{- {1\over 2} \int d^3k \int d^3k^\prime \tilde{\phi}_k 
\tilde{\phi}_{k'}}  \, . \label{OOpathint}
\end{equation}
Now that we have disentangled the path integral from the correlation
function it should be clear that the path integral is going to give
just a number.  In that case this leaves us with the following
relation between bulk expectation values and ``boundary'' correlators,
which was first derived in~\cite{Maldacena},
\begin{equation}
\left< \phi_k \phi_{-k} \right> \propto {-1 \over 2 {\rm Re} \, \left< 
{\cal O}_k {\cal O}_{-k} \right>} \, . 
\label{expv-2ptfn}
\end{equation}
We note that because the bulk scalar field acts as a source in the
CFT, this relation is almost like a Legendre transform, except for the
factor of two and the restriction to the real part of the correlators,
which can respectively be traced back to the square and the norm of
the scalar field wave functional in~(\ref{bulkev}).

\subsection{CFT operators, energy momentum and inflationary perturbations}

Next we want to explicitly show that one obtains the correct
expressions for the spectrum of inflationary perturbations, starting
from CFT correlation functions. The vacuum 2-pt function in three
euclidean dimensions of operators with arbitrary scaling dimensions
$\Delta$ is given by
\begin{equation}
\left< {\cal O}_{\Delta}(x) {\cal O}_{\Delta}(x^\prime) \right> \propto 
{ \eta_c^{2\Delta - 6} \over |x-x^\prime|^{2\Delta}} H^{-2} \, , 
\label{cftcorx}
\end{equation}
where $\eta_c$ corresponds to a cutoff length scale necessary to
properly define the correlator. After Fourier transforming this reads
\begin{equation}
\left< {\cal O}_{\Delta} {\cal O}_{\Delta} \right> \propto 
H^{-2} |k|^3 \left( { \mu_c \over |k|} \right)^{6 - 2\Delta} , 
\label{cftcork}
\end{equation}
where $\mu_c = 1/\eta_c$ now corresponds to an energy cutoff. Note
that the explicit cutoff dependence disappears when one considers
marginal operators $\Delta=3$. We will write $k=|k|$ from now on.

We will use the expression~(\ref{scpower}) together
with~(\ref{sfpower}) to calculate the power spectrum of inflationary
density perturbations.  The appearance of the function $\beta$ in
these expressions has a beautiful interpretation in the CFT. First
note that the bulk metric acts as the source for the energy momentum
tensor in the CFT~(\ref{EMTsource}).  It can be shown that to discuss
scalar curvature perturbations we should be interested in the trace of
the energy momentum tensor~\cite{Maldacena}. In a CFT the expectation
value of the trace of the energy momentum tensor vanishes. However,
moving away from the conformal fixed point one finds the well known
and intuitively understood expression for the conformal anomaly in
terms of non-marginal CFT operators (we are only considering flat
spatial backgrounds)
\begin{equation}
\left< {\rm Tr} \, T^{ij} \right> = \beta(g) \, {\partial \over 
\partial g} {\cal L} = \beta(g) \, \left< {\cal O}_\Delta 
\right> , \label{confanomaly}
\end{equation}
where $\beta(g)$ is the beta-function for the coupling $g$ appearing
in the lagrangian deforming the CFT~(\ref{cftdeform}), i.e.\ ${\cal L}
= g \, {\cal O}_\Delta$. So we see the beta-function appearing
naturally. Realizing that the expectation value of density
perturbations would be proportional to the inverse of the 2-pt
correlation function of the trace of the energy momentum tensor we
reproduce the $\beta^{-2}$ factor in~(\ref{scpower}). So in terms of
CFT correlators the power spectrum of density fluctuations can be
written as
\begin{equation}
{\cal P}_{\cal R} \propto k^3 {1 \over {\rm Re} \left< T^i_i \, T^j_j
\right>} = k^3 {1 \over \beta^2 \, {\rm Re}  
\left< {\cal O}_\Delta {\cal O}_\Delta \right>} \, . \label{power-cftcor}
\end{equation}
From the general expression for the CFT correlator~(\ref{cftcork}) it
seems that we will get explicit dependence on the cutoff $\mu_c$ in
the power spectrum. However, one should realize that $\beta(g)$ also
depends on the cutoff $\mu_c$ as follows
\begin{equation}
\beta(g) \approx (\Delta -3)\, g + \cdots = 
(\Delta -3) \, \mu_c^{\Delta-3} + \cdots \, . 
\label{betapprox}
\end{equation}
This dependence will exactly cancel the cutoff dependence of the
correlation function, giving us a cutoff independent answer, as was
also discussed in section 3.2. Note that this is also consistent with
the fact that the energy momentum tensor in the bulk should correspond
to a marginal operator in the CFT and should therefore not depend on
the cutoff. The final result is
\begin{equation}
{\cal P}_{\cal R} \propto k^3 {1 \over \beta^2 \, {\rm Re}  
\left< {\cal O}_\Delta {\cal O}_\Delta \right>} = H^2 \, k^{6 - 2\Delta}
= H^2 \, k^{-2\lambda} \, , \label{cft-kdep}
\end{equation}
where $\lambda=\Delta -3 = -3/2 \pm \sqrt{9/4 - m^2/H^2}$. So we
reproduce the expected result for the spectral index to lowest order
in the departure from the fixed point~(\ref{indexexact}) corresponding
to the limit $g \rightarrow 0$, corresponding to $\beta= 0$
in~(\ref{indexexact}).

\subsection{RG-flow equations and the spectrum of inflationary perturbations}

Let us now consider adding the following deformation to the CFT action
\begin{equation}
{\cal S} = {\cal S}_{\rm CFT} + g \int d^3x {\cal O}_{\Delta} \, .
\label{CFTdeform}
\end{equation}
This deformation will break conformal invariance and lead to RG-flow.
The CFT correlators that were introduced in the previous section will
now change along the RG-flow trajectory. This will therefore affect
the spectrum of inflationary density perturbations as one moves
further away from the conformal fixed point. To systematically follow
this process one needs to study the RG-flow equations, in particular
of course the Callan-Symanzik equations for the
correlators. In~\cite{Larsen-McNees} the RG-flow equations were
carefully deduced from the bulk dS gravity theory (see
also~\cite{NO}), following an approach first applied in the AdS/CFT
context by~\cite{BVV}. Introducing the Hamilton-Jacobi action
functional in the bulk one carefully introduces counterterms for the
local divergences that appear in the limit $t \rightarrow \infty$,
defining an effective renormalized action. This new action functional
also satisfies the Hamilton-Jacobi equation that after some rewritings
and neglecting of higher order terms can be viewed as an RG equation
for the renormalized action. For all the details we refer
to~\cite{Larsen-McNees, NO}. We will be interested in the evolution of
the correlators, in particular the $k$ dependence along the flow that
will determine the spectral index of the inflationary density
perturbations.

As already explained Callan-Symanzik equations can be deduced from the
bulk dS gravity theory. We will be interested in 2-pt functions
$\left< {\cal O}(x) \, {\cal O}(y) \right>$ only, and in that case the
Callan-Symanzik equation looks as follows
\begin{equation}
\left[  \Lambda {\partial \over \partial \Lambda} + \beta(g) {\partial \over
\partial g} + 2 \lambda(g) \right] \left< {\cal O}_\Delta {\cal O}_\Delta
\right> = 0 \, , \label{CSa-eqn}
\end{equation}
where $\beta(g) \equiv \Lambda {\partial g \over \partial \Lambda}$
with $\Lambda = \mu_c$ corresponding to the cutoff scale.  The
(anomalous) scaling dimension $\lambda(g)$ is defined as follows
\begin{equation}
\lambda (g) \equiv {\partial \beta \over \partial g} = \lambda + \eta(g) 
= \Delta-3 + \eta(g) \, , \label{anscaling}
\end{equation}
where $\lambda$ and $\Delta$ correspond to the constant scaling
dimensions at a conformal fixed point (i.e.\ the lowest order
coefficients in an expansion of $\beta(g)$ in $g$).  Even though this
definition of the scaling dimension is not the standard one, it
reproduces the standard definition of non-marginal scaling dimensions
in the limit of small coupling $g$. This can be seen by first noting
that $g$ can be identified with the field-strength renormalization
factor $g=Z^{1/2}$ because of the particular
deformation~(\ref{CFTdeform}) we are considering. Then using the more
standard definition of the anomalous scaling dimension $2 \lambda(g) =
{\Lambda \over Z} {\partial Z \over \partial \Lambda} = {\beta \over
  g}$, we see that non-marginal scaling dimensions $\lambda \neq 0$ in
the CS-equation are reproduced as the lowest order coefficients in an
expansion in the coupling $g$. The higher order coefficients $a_i$,
$i>1$, are related by a factor of $i$ between the two definitions.

As a consequence of straightforward dimensional analysis one can
determine the cutoff dependence of a general Fourier transformed
correlator as
\begin{equation}
\left< {\cal O}_\Delta {\cal O}_\Delta \right> (k) = k^3 F \left[ 
{\Lambda^2 \over k^2} , g(\Lambda) \right] , \label{momcor}
\end{equation}
which is just the generalized version of the
expression~(\ref{cftcork}) with $\Lambda = \mu_c$.  Using this
expression it is possible to rewrite the CS-equation in terms of a
momentum $k$-derivative instead of the cutoff scale $\Lambda$ by
noting that
\begin{equation}
\Lambda {\partial \over \partial \Lambda} \left< {\cal O}_\Delta 
{\cal O}_\Delta \right> 
(k) =  \left( -k {\partial \over \partial k} + 3 \right)\, \left< {\cal O} 
{\cal O} \right> (k) \, . \label{lambda-k}
\end{equation}  
This allows us to rewrite the CS-equation~(\ref{CSa-eqn}) as follows
\begin{equation}
\left[ {\partial \over \partial {\rm ln}k} - \beta(g) {\partial \over
\partial g} + (3-2\Delta) - 2\eta(g)) \right] \left< {\cal O}_\Delta 
{\cal O}_\Delta \right> = 0 \, . \label{CS-eqn}
\end{equation}
The $k$-dependence of the CFT 2-pt correlators~(\ref{cftcork}) is
reproduced in the fixed point limit $g \rightarrow 0$, i.e.\ when one
recovers exact conformal invariance so $\beta(g)=\eta(g)=0$.  We will
be interested in corrections to the $k$-dependence of the correlator
due to RG-flow away from the fixed point.  There is a quick and dirty
way to see that one can reproduce the approximate expression for the
spectral index~(\ref{avpowerspec}).  It first of all involves the
assumption that the CFT correlators~(\ref{cftcorx}) only depend on the
coupling $g$ through the Hubble parameter $H$, which as we will see
later on is not expected to be true in general.  Nevertheless
proceeding, using the expression~(\ref{Hbeta}) for the beta-function
and remembering that the coupling $g$ corresponds to the scalar field
in the bulk, we derive that
\begin{equation}
\beta(g) {\partial \over \partial g} \left< {\cal O}_\Delta {\cal O}_\Delta
\right> = \beta^2 \left< {\cal O}_\Delta {\cal O}_\Delta \right> . 
\label{dgcor}
\end{equation}
This relation allows one to solve the Callan-Symanzik
equation~(\ref{CS-eqn}) if one also assumes an approximately constant
beta-function $\beta(g) \approx \bar{\beta}$ and scaling dimension
$\lambda(g) \approx \bar{\lambda} = \bar{\Delta} -3 = \Delta -3 +
\bar{\eta}$. In that case we find that the $k$ dependence of the
correlator will be modified to become
\begin{equation}
\left< {\cal O}_\Delta {\cal O}_\Delta \right> \propto k^3 \, 
k^{\bar{\beta}^2 + 2\bar{\Delta} -6} \, , 
\label{betamodk}
\end{equation}
which will indeed reproduce the expression~(\ref{avpowerspec}) for the
spectral index of the power spectrum of density perturbations,
using~(\ref{power-cftcor}).  The condition of approximately constant
$\bar{\beta}$ and $\bar{\lambda}$ can be made more precise by saying
that ${\partial \over \partial \ln k} \beta = \beta \lambda \ll 1$ and
${\partial \over \partial \ln k} \lambda = \beta {\partial \over
  \partial g} \lambda \ll 1$, so typically these conditions can both
be satisfied when $\beta \ll 1$. Although we reproduce the correct
answer this way, this approach is not completely transparent and it
will be useful, especially if we want to consider further corrections,
to develop a more consistent method toward solving the CS-equation.

As is well-known, a more general approach toward solving the
CS-equation would be to start with the following ansatz for the 2-pt
correlator
\begin{equation}
\left< {\cal O}_\Delta {\cal O}_\Delta \right> \propto Z(g) k^{3+2\lambda} \, ,
\label{CSansatz}
\end{equation}
separating the full $k$ and $g$ dependence. To solve the CS-equation
the function $Z(g)$ is expressed as
\begin{equation}
Z(g) = Z_0 \, \exp{\int_0^g dg' {2\lambda - 2 \lambda(g') \over
\beta(g')}} = Z_0 \exp{-2 \int_{\Lambda_0}^{\Lambda_g} d\ln
\left(\frac{\Lambda'}{k}\right) \eta(\Lambda')} \, . 
\label{Zint}
\end{equation}
One can easily check that this expression for $Z(g)$ solves the
CS-equation.  The expression in terms of an integral over
$d\ln(\Lambda/k)$, with $k$ introduced as the only other natural scale
that can appear in the correlator, has the advantage of making it
possible to directly read off the $k$-dependence that we are
interested in. The only thing left to do is to evaluate the function
$Z(g)$ (or $Z(\Lambda/k)$), which for example can be done order by
order in $g$. One can in fact similarly express the Hubble parameter
$H^{-2}$ in terms of an integral over the coupling (or the logarithmic
cutoff scale)~(\ref{Hbeta}),~(\ref{cintN})
\begin{equation}
H(g)^{-2} = H_0^{-2} \exp{\int_0^g dg' \beta(g')} = H_0^{-2} \exp{
\int_{\Lambda_0}^{\Lambda_g} d\ln\left(\frac{\Lambda'}{k}\right) 
\beta(\Lambda')^2} \, . \label{Hintg}
\end{equation}
This allows us to write the solution $Z(g)$ as follows 
\begin{eqnarray}
Z(g) &=& H(g)^{-2} \, \exp{\int_0^g dg' {- 2 \eta(g') -\beta(g')^2 \over
\beta(g')}} \\ \nonumber 
&=& H(g)^{-2} \exp{\int_{\Lambda_0}^{\Lambda_g} 
d\ln\left(\frac{\Lambda'}{k}\right) \left( -2 \eta(\Lambda') - 
\beta(\Lambda')^2 \right) } \, , \label{Z-H}
\end{eqnarray}
extracting the $H^{-2}$ part from the solution $Z(g)$, with $\eta(g)$
corresponding to the anomalous part of the scaling
dimension~(\ref{anscaling}). From this general expression it is more
transparent under precisely what conditions we can reproduce the
spectral index of inflationary perturbations, and how to go beyond
that.  Isolating the Hubble parameter in $Z(g)$ means that we are
interested in the induced $k$-dependence relative to the average
Hubble parameter, which is exactly what is done in the bulk
calculation of the inflationary power spectrum. The integral in the
second line of~(\ref{Z-H}) can be approximated to lowest order by
assuming constant $\beta=\bar{\beta}$ and $\eta
=\bar{\eta}$. Using~(\ref{power-cftcor}) this immediately reproduces
the expression for the spectral index of the scalar curvature
perturbation we found earlier through a more ``sloppy'' method,
agreeing with the bulk result~(\ref{avpowerspec}). However, starting
from~(\ref{Z-H}) would in principle allow us to go beyond the
approximation of constant $\beta$ and $\lambda$~\cite{Kinney}. We hope
to come back to that problem in the near future.
  
\section{Holography and universality of inflationary perturbations}

Just as in AdS/CFT, in dS/CFT the UV and IR are interchanged with
respect to bulk or boundary physics. A UV cutoff $\Lambda$ in
euclidean field theory corresponds to an IR (large time/scale factor)
cutoff $t_c$ in the bulk.  The UV cutoff in the field theory implies
that we will only be considering euclidean momenta $|k| < \Lambda$ and
realizing that $|k|$ corresponds to the comoving momentum and that the
cutoff can be written as $\Lambda= a_c H$ in the bulk we find the
following constraint on the bulk physical momentum $|p| = |k|/a$
\begin{equation}
p \leq H \, . \label{physcon}
\end{equation}
The fact that this constraint on the physical momentum in the bulk
only depends on the Hubble parameter tells us that dS/CFT only probes
bulk physics outside the de Sitter horizon. Bulk physics inside the de
Sitter horizon, as experienced by a free-falling observer, is
inaccessible because it will always be integrated out in the CFT
description.  In terms of coarse graining it means that a static de
Sitter patch always corresponds to a single lattice site in the field
theory, no matter what the size of the cutoff scale in the field
theory. One can in fact naturally associate an entropy to a single
coarse grained region that is roughly equal to the central charge of
the CFT, which is reproducing the de Sitter entropy that can be
associated to a static patch of de Sitter~\cite{Kabat}, because the
central charge indeed scales like the area of the static patch Hubble
volume~(\ref{S-c}). However, the crucial reason for the entropy to
scale with the area of one Hubble volume is because it corresponds to
a single lattice site in the field theory. As soon as one considers
larger physical bulk volumes containing more than a single lattice
site one will find that the entropy will scale with the total
three-dimensional volume, instead of the area, as expected from a
three-dimensional field theory.  What dS/CFT does for us is to
identify a single Hubble volume as the unique volume at which the
entropy indeed scales like the area, corresponding to a single lattice
site in the field theory. This is all dS/CFT will ever have to say
about holography in the static patch, because physics inside the
static patch can not be probed by dS/CFT. One can therefore question
the validity of objections raised against dS/CFT that rely on physics
in the de Sitter static patch~\cite{susskind}. The applicability, if
any, of dS/CFT should instead be on bulk length scales larger than the
Hubble radius.  This is in fact just right for a treatment of the
spectrum of inflationary perturbations using CFT correlators, which is
only formed after wavelengths cross the Hubble radius. This is of
course what we have shown, that one can calculate the spectrum of
inflationary perturbations using deformed CFT correlators.

In the previous paragraph we pointed out that, according to dS/CFT,
the holographic area-entropy relation immediately breaks down on
length scales larger than the Hubble radius and the ordinary scaling
of entropy with the total volume is found instead. We therefore
conclude that even though dS/CFT is considered to be a ``holographic''
correspondence, it does not predict any holographic constraints
(having to do with the area law for the gravitational entropy) on the
spectrum of inflationary perturbations.  Similar ideas in support of
the (near) absence of holographic constraints on the CMB spectrum have
been described in~\cite{UD-hol}, however opposite claims have also
been reported~\cite{CMB-Holcon}. It would perhaps be interesting to
study this in more detail. This slightly counter-intuitive result is
due to the fact that in dS/CFT the ``holographic'' coordinate is
timelike and holography therefore acquires a completely different
meaning, as we will now discuss.

The usual bulk understanding of inflationary perturbations is as
originating from vacuum fluctuations that grow to become the size of
the Hubble horizon after which they freeze and become part of the
spectrum of inflationary perturbations. This understanding does not
immediately explain why the spectrum of inflationary perturbations
seems to be so universal, i.e.\ very independent of microscopic
physics. In this new approach based on deformed CFT correlators and
RG-flow equations the static patch is completely removed (or coarse
grained) and nevertheless we still recover the correct spectrum of
inflationary perturbations. As already explained the static patch,
corresponding to a single Hubble volume, corresponds to scales above
the cutoff from the deformed CFT point of view. This necessarily
implies that the spectrum of inflationary perturbations is very
independent of the physics at bulk (length) scales smaller than the
horizon, except for its effect on certain universal parameters
(homogeneous fields in the bulk and couplings in the deformed CFT), in
exactly the spirit of the Wilsonian renormalization group.  It is this
concept of universality that should be the main message of (timelike)
holography. The deformed CFT approach explicitly realizes the strong
independence of the spectrum of inflationary perturbations from
detailed sub-horizon or trans-Planckian dynamics. So one advantage of
the CFT approach is that it makes the universal nature of the
primordial CMB spectrum explicit. Note that the primordial CMB
spectrum can and will of course still depend on initial (boundary)
conditions that one imposes on sub- and/or super-horizon scales. In
the quantum theory this corresponds to selecting a different vacuum
which will affect the spectrum, a fact that has attracted a lot of
attention recently~\cite{Alpha}.  From the CFT point of view, changing
boundary conditions or vacua in the bulk will affect the bulk action
and therefore the CFT correlators, see also~\cite{dS/CFT-vacua}. It
would perhaps be interesting to study this in more detail. We hope to
come back to these and other matters in the near future.

\section{Note added}

During the final stages of this work~\cite{Larsen-McNees} appeared in
which essentially the same ideas are described. In particular our
paper has a lot of overlap with section 6 in their paper. We have
tried to highlight the dS/CFT point of view and commented on
holography and universality in the context of the power spectrum of
inflationary perturbations. Instead of our emphasis on a UV field
theoretic point of view, the authors of~\cite{Larsen-McNees} focus
more on the complementary bulk gravitational IR point of view, which
is very interesting in its own right.

\acknowledgments

I would like to thank F. Larsen, R. Leigh, D. Schwarz, D. Chung,
A. Petkou, D. Kabat, D. Minic, K. Schalm and G. Shiu for many
interesting and useful discussions. I thank K. Schalm for reading a
first draft and suggesting improvements. The Amsterdam Summer Workshop
is acknowledged for hospitality and for providing an inspiring
atmosphere.

\end{document}